\documentclass{article}

\usepackage{amssymb,amsfonts,amsmath,stmaryrd}
\usepackage{cite,enumerate,float,indentfirst}
\usepackage{color}

\def\be{\begin{eqnarray}}
\def\ee{\end{eqnarray}}
\def\nn{\nonumber}

\def\p{\partial}

\def\Z{x}
\def\L{y}

\newcommand{\beq}{\begin{equation}}
\newcommand{\eeq}{\end{equation}}
\newcommand{\beqa}{\begin{eqnarray}}
\newcommand{\eeqa}{\end{eqnarray}}

\definecolor{red}{rgb}{1,0,0}
\definecolor{orange}{rgb}{1,0.5,0}
\definecolor{violet}{rgb}{0.7,0,1}

\newcommand\br[1]{
  \left(#1\right)
}

\voffset= - 1.1in
\hoffset= - 1.0in         

\begin{document}

\title{\vspace{1.5cm}\bf
Chalykh's Baker-Akhiezer functions as eigenfunctions\\
of the integer-ray integrable systems
}

\author{A. Mironov$^{b,c,d,}$\footnote{mironov@lpi.ru,mironov@itep.ru},
A. Morozov$^{a,c,d,}$\footnote{morozov@itep.ru},
A. Popolitov$^{a,c,d,}$\footnote{popolit@gmail.com}
}

\date{ }

\maketitle

\vspace{-6cm}

\begin{center}
  \hfill MIPT/TH-24/24\\
  \hfill FIAN/TD-15/24\\
  \hfill ITEP/TH-30/24\\
  \hfill IITP/TH-25/24
\end{center}

\vspace{4.5cm}

\begin{center}
$^a$ {\small {\it MIPT, Dolgoprudny, 141701, Russia}}\\
$^b$ {\small {\it Lebedev Physics Institute, Moscow 119991, Russia}}\\
$^c$ {\small {\it NRC ``Kurchatov Institute", 123182, Moscow, Russia}}\\
$^d$ {\small {\it Institute for Information Transmission Problems, Moscow 127994, Russia}}
\end{center}

\vspace{.1cm}

\begin{abstract}
Macdonald symmetric polynomial at $t=q^{-m}$ reduces to a sum
of much simpler complementary {\bf non-symmetric} polynomials,
which satisfy a simple system of the {\bf first order} linear difference equations with constant coefficients,
much simpler than those induced by the usual Ruijsenaars Hamiltonians of the cut-and-join type.
We provide examples of explicit expressions for these polynomials
nicknamed Baker-Akhiezer functions (BAF), and demonstrate that they further decompose into
sums of {\bf nicely factorized} quantities, perhaps, non-uniquely.
Equations and solutions can be easily continued to {\bf non-integer} parameters $\lambda$,
which, in Macdonald polynomial case, are associated with integer partitions.
Moreover, there is a straightforward generalization to ``twisted" BAF's,
which, however, are not so easy to decompose,
and factorization of the coefficients is lost, at least naively.
Still, these twisted BAF's provide eigenfunctions for Hamiltonians
associated with commutative integer ray subalgebras of the Ding-Iohara-Miki algebra.
\end{abstract}

\bigskip

\newcommand\smallpar[1]{
  \noindent $\bullet$ \textbf{#1}
}

\section{Introduction}

Recently in \cite{MMP1} we managed to include into the modern context of representation theoretic
structure of the Ding-Iohara-Miki (DIM) algebra an interesting approach\footnote{This approach is basically a generalization of an earlier construction \cite{CV,Ves,Cha2} of eigenfunctions of the rational and trigonometric Calogero-Sutherland systems to the Ruijsenaars case.} by O. Chalykh to description of the Macdonald polynomials \cite{Cha} (see also an earlier work \cite{ES}).
O. Chalykh proposed to describe the Macdonald polynomials $M_{\vec\lambda}[\vec x]$ \cite{Mac}
as a lifting (analytical continuation) in $t$ from the points $t=q^{-m}$ with integer non-negative $m$,
where they reduce to a sum of much simpler polynomials $\Psi_m[\vec z,\vec\lambda]$ in $x_i=q^{z_i}$,
related by action of the Weyl group. These polynomials are
nicknamed multivariable Baker-Akhiezer functions (BAF's) after \cite{CFV}. The main advantage of the Chalykh's approach
is that {\bf the BAF's satisfy linear difference equations with constant coefficients}.
These equations are much simpler that the conventional ones, made from the Ruijsenaars finite-
difference version of cut-and-join operators.\footnote{There are many papers contributed in various directions nearby, up to the theory of multiple integrals of the Mellin-Barnes type in \cite{Kharchev}, and to the cut-and-join operators in \cite{MMN}. Here our direction is, however, quite orthogonal and reveals very different sides of the story.}
What is even more important, the BAF equations allow for a new deformation, to ``twisted" BAF's,
which depend on one more positive integer parameter $a$.
Then the solutions at $a>1$ are no longer related to the Macdonald polynomials,
but instead to some more involved symmetric polynomials.
In terms of integrable Hamiltonians, this substitutes the Ruijsenaars operators by the ``higher-ray" DIM Hamiltonians $\hat H^{(-1,a)}_k$ introduced in \cite{MMP} within the studies of q,t-matrix models \cite{Ch3} (see also previous works \cite{MPSh,Max,MMell}) generalizing the WLZZ models \cite{China,Ch12} along the line of \cite{MMMP12}.
It turns out that {\bf the twisted BAF's provide eigenfunctions of these higher $\hat H^{(-1,a)}_k$  for integer rays}.
Actually this was an original observation by O. Chalykh and M. Fairon \cite{ChF} for the Hamiltonian $\hat H^{(-1,2)}_1$, much before the meaning and generalization
of this  operator was revealed in \cite{MMP}, and it was Chalykh's educated guess afterwards
that this would work for all $a\geq 2$, as has been now partly confirmed in \cite{MMP1}.

\bigskip

The goal of the present paper is to provide some explicit examples for somewhat sophisticated presentation
in \cite{Cha,ChE,MMP1}, to demonstrate that the calculations are indeed constructive and relatively straightforward.
Still we did not yet manage to make them in full generality
(at least, preserving the relative simplicity), and to obtain generic formulas for the twisted BAF's
$\Psi^{(a)}_m[\vec z,\vec\lambda]$, which provide common eigenfunctions for all the commuting Hamiltonians $\hat H^{(-1,a)}_k$
for a given ray with parameter $a$
(Hamiltonians belonging to this ray are enumerated by $k$, and eigenfunctions, by $\vec\lambda$).
Nor did we find eigenfunctions for rational rays $\hat H^{(-b,a)}_k$ with arbitrary coprime $a$ and $b$.
All this requires deeper insights and remains for the future study by the community.

\bigskip

In this paper, we deal with the simplest examples in three directions:
\begin{itemize}
\item{} Non-twisted $\Psi^{(1)}$ for $N>2$.
Here $N$ is the number of variables $x_i$.

\item{} Examples of $\Psi^{(a)}$ for $a>1$ and $N=2$.
  These are already somewhat non-trivial, and their closed form,
  and the structures involved, remains a question.
  The increase of $N$, while straightforward,
  appears to be too technical and not illuminating.

\item{} Examples of how these $\Psi^{(a)}_m$ defined as solutions to Chalykh's linear equations \eqref{Psideco},\eqref{linsys0}
  and \eqref{eq:twba-ansatz},\eqref{linsys}
become eigenfunctions of appropriately twisted Hamiltonians  $\hat H^{(-1,a)}_k$.

\end{itemize}

Devoted to these three directions are respectively
sec.\ref{sec:solving-linear}-\ref{genarg},
sec.\ref{twBAF1}-\ref{twBAF2} and sec.\ref{Hams}.
We end in sec.\ref{conc} with a short conclusion.

\paragraph{Notation.} Throughout the text, we use the notation for the variables $x_i=q^{z_i}$ and $y_i=q^{\lambda_i}$ for the BAF depending on the variables $\vec z=(z_1,z_2,\ldots,z_N)$ and $\vec\lambda=(\lambda_1,\lambda_2,\ldots,\lambda_N)$.

The $q$-Pochhammer symbol is defined as
\be
(w;q)_n=\prod_{i=0}^{n-1}(1-wq^i)
\ee

\section{Macdonald polynomials and BAF\label{redMac}}

Before proceeding with BAF's,
we once again stress {\it a striking difference} of BAF from Macdonald polynomials,
which inspired their original discovery
(related also to the CMM theory \cite{MM-conj,Ch1,EK}, see also \cite{MMPCMM}):
BAF's satisfy {\it first-order} difference equations and are defined for {\it arbitrary} complex
values of $\lambda_i$, which, in the Macdonald case, were associated with integer partitions.
More concretely, the relation between the Macdonald polynomial and the BAF is
\be\label{MPsi}
M_{\vec\lambda+m\vec\rho}(\{q^{z_i}\};q,q^{-m})=
{\cal N}_\lambda^{-1}\cdot\sum_{w\in W}\Psi_m(w\vec z,\vec\lambda)
\ee
Here $w$ is an element of the Weyl group of $A_{N-1}$, $\vec\rho$ is the Weyl vector, and $\mu:=\vec\lambda+m\vec\rho$ is an integer partition, $\mu_1\ge\mu_2\ge\ldots\mu_N\ge 0$.

Note that the entire theory of BAF deals non-symmetric (quasi)polynomials, hence, it does not admit a formulation in terms of power sums $p_k=\sum_i^Nx_i^k$, which is quite effective in the case of Macdonald polynomials. However, dealing with the Macdonald polynomial as a symmetric polynomial of $x_i$ has one important advantage: one can use the decomposition
\be
M_\mu[\vec x'+\vec x''] = \sum_\nu M_{\mu/\nu}[\vec x'']M_\nu[\vec x']
\ee
where $M_{\mu/\nu}[\vec x]$ denotes the skew Macdonald polynomial and choose $\vec x'=(x_1,x_2,\ldots,x_{N-1},0)$, $\vec x''=(0,0,\ldots,0,x_N)$. Then, one gets \cite[Eqs.(6.20),(6.24),(7.14')]{Mac}
\be
M_\mu[\vec x] = \sum_\nu x_N^{|\mu|-|\nu|}M_{\mu/\nu}[1]M_\nu[\vec x']
\ee
with an explicit formula for $M_{\mu/\nu}[1]$, which is non-zero only if $\mu-\nu$ is a horizontal strip.
Now repeating this procedure one obtains the Macdonald polynomial as a sum \cite{NS}
\be
M_\mu[\vec x] =\prod_{i=1}^Nx_i^{\lambda_i}\sum_{\{\mu^{j}\}}\left(\prod_{k=1}^NM_{\mu^{(k)}/\mu^{(k-1)}}[1]\right)\prod_{1\le i<j\le N}
\left({x_j\over x_i}\right)^{\mu_i^{(j)}-\mu_i^{(j-1)}}
\ee
with $\mu^{(N)}=\mu$.

This formula implies a strong asymmetry of the ``elementary" building blocks in $x_i$.
At the same time, in \cite{NS} there was found a representation for $M_{\mu/\nu}[1]$,
\be
M_{\mu/\nu}[1]=\prod_{1\le i<j\le N}{(q^{\nu_i-\mu_j+1}t^{j-i-1};q)_{\mu_i-\nu_i}\over(q^{\nu_i-\mu_j}t^{j-i};q)_{\mu_i-\nu_i}}
\cdot\prod_{1\le i\le j\le N}{(q^{\nu_i-\mu_j+1}t^{j-i-1};q)_{\mu_i-\nu_i}\over(q^{\nu_i-\mu_j}t^{j-i};q)_{\mu_i-\nu_i}}
\ee
when each block can be easily continued from integer to arbitrary complex values of $\vec\mu$.
The only point is that for such $\vec\mu$ the sums are no-longer cut by the Pochhammer factors,
and polynomials become series.
This makes the emergent BAF's a very broad continuation of Macdonald polynomials
actually to the Noumi-Shiraishi functions \cite{NS}, which we consider in the role of BAF's
in a forthcoming paper \cite{MMP3}.

\section{Solving defining linear equations for non-twisted BAF
\label{sec:solving-linear}}

Following \cite{Cha} (see also \cite{MMP1}), in order to construct the BAF,
one needs to determine the coefficients $\psi$ in the expansion
\be
\Psi_m(\vec z,\vec\lambda)
= q^{(\vec\lambda+m\vec\rho)\vec z} \sum_{k_{ij}=0}^m q^{-\sum_{i>j}k_{ij}(z_i-z_j)}\psi_{m,\vec\lambda,\{k\}}
\label{Psideco}
\ee
from the {\it linear} conditions
\be
\Psi_m(z_k+j,\vec\lambda) =\Psi_m(z_l+j,\vec\lambda), \ \ \ 1\leq j \leq m
\label{linsys0}
\ee
imposed for any pair of $k,l$ at arbitrary values of $\vec z$, constrained by single requirement
that $\Z_k=\Z_l$ or $q^{z_k}=q^{z_l}$.

In this section we consider   different $N$, therefore we use the notation $\Psi^{[N]}_m$ with superscript $[N]$
in square brackets.
Hopefully this will not cause confusion with the other sections, where $N=2$ and the superscript is $(a)$.
In this section $a=1$, and it is suppressed.
Also we suppress $\lambda$ to make the formulas more readable
and, where appropriate, use logarithmic notation $\L_i = q^{\lambda_i}$.

\section*{$N=2$}

In this case there is no ambiguity in the solution
\be
\Psi^{[2]}_m(x_1,x_2) = x_1^{\lambda_1}x_2^{\lambda_2}\sum_{k=0}^m \left(\frac{\Z_1}{\Z_2}\right)^{\frac{m}{2}-k} \psi^{(1,2)}_{m|k}
\ee
with
\be
\boxed{
\psi^{(A,B)}_{m|k}:= \frac{[m]!}{[k]![m-k]!} \prod_{l=1}^k \frac{q^{m-l+1}\L_A-\L_B}{q^{l}\L_B-\L_A}
}
\label{psiAB}
\ee
where $[n]:=\frac{q^{-n}-1}{q^{-1}-1}$. In this section, for the sake of brevity, we omit the arguments $\vec\lambda$ of BAF.
Then for any $j = 1,\ldots, m$
\be
\left(\frac{\L_1}{\L_2}\right)^j \frac{\Psi^{[2]}_m(\Z_1q^j,\Z_1)}{\Psi^{[2]}_m(\Z_1,\Z_1q^j)} = 1
\ee
These equations for $N=2$ do not actually depend on $\Z_1$.

\section*{$N=3$}

This case is already fully non-trivial.
We have
\be \label{eq:n3psi-ansatz}
\Psi^{[3]}_m(x_1,x_2,x_3) =\prod_{i=1}^3x_i^{\lambda_i} \cdot
\left(\frac{\Z_1}{\Z_3}\right)^m \sum_{k_{12},k_{23},k_{13}=0}^m
 \left(\frac{\Z_1}{\Z_2}\right)^{-k_{12}} \left(\frac{\Z_2}{\Z_3}\right)^{-k_{23}} \left(\frac{\Z_1}{\Z_3}\right)^{-k_{13}}
 \underbrace{\psi^{(1,2)}_{m|k_{12}}  \psi^{(2,3)}_{m|k_{23}}  \psi^{(1,3)}_{m|k_{13}}}_{
   \substack{\text{standard factors \eqref{psiAB}} \\ \text{as before}}
 }\cdot  \underbrace{\tilde\psi_{m|k_{12},k_{23},k_{13}}}_{\text{new contribution}}
\ee
subjected to three sets of equations:
\be
\L_1^j \Psi^{[3]}_m(\Z_1q^j,\Z_1,\Z_3)=\L_2^j\Psi^{[3]}_m(\Z_1,\Z_1q^j,\Z_3) \nn \\
\L_1^j \Psi^{[3]}_m(\Z_1q^j,\Z_2,\Z_1)=\L_3^j\Psi^{[3]}_m(\Z_1,\Z_2,\Z_1q^j) \nn \\
\L_2^j \Psi^{[3]}_m(\Z_1,\Z_2q^j,\Z_2)=\L_3^j\Psi^{[3]}_m(\Z_1,\Z_2,\Z_2q^j),
\label{eqs3}
\ee
where $j$ runs from $1$ to $m$, and the values of $\Z_i$ are arbitrary.
Actually, these are polynomials in $\Z_i$, and therefore each line in \eqref{eqs3} is
a finite  (but large) collection of conditions
on the $\Z$-independent coefficients, split for convenience into
already familiar $\psi$ and new quantities $\tilde\psi$.

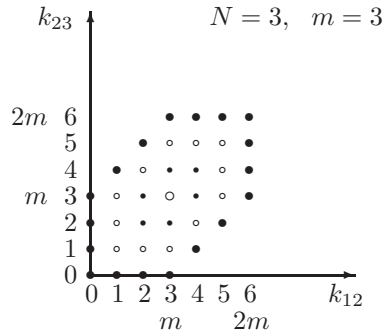
\begin{figure}[h]
\begin{picture}(150,150)(-150,-30)

\put(0,0){\vector(1,0){100}}
\put(0,0){\vector(0,1){100}}
\put(0,0){\circle*{3}} \put(10,0){\circle*{3}}\put(20,0){\circle*{3}} \put(30,0){\circle*{3}}
\put(0,10){\circle*{3}}\put(0,20){\circle*{3}}\put(0,30){\circle*{3}}
\put(40,10){\circle*{3}}\put(50,20){\circle*{3}} \put(60,30){\circle*{3}}
\put(10,40){\circle*{3}}\put(20,50){\circle*{3}}\put(30,60){\circle*{3}}
\put(40,60){\circle*{3}}\put(50,60){\circle*{3}}\put(60,60){\circle*{3}} \put(60,40){\circle*{3}}\put(60,50){\circle*{3}}

\put(10,10){\circle{2}} \put(10,20){\circle{2}}\put(10,30){\circle{2}} \put(20,10){\circle{2}} \put(30,10){\circle{2}}
\put(20,40){\circle{2}} \put(40,20){\circle{2}}
\put(50,50){\circle{2}} \put(50,40){\circle{2}}\put(50,30){\circle{2}} \put(40,50){\circle{2}} \put(30,50){\circle{2}}

\put(20,20){\circle*{2}} \put(30,20){\circle*{2}}\put(20,30){\circle*{2}}
\put(40,40){\circle*{2}} \put(30,40){\circle*{2}}\put(40,30){\circle*{2}}

\put(30,30){\circle{3}}

\put(90,-10){\mbox{$k_{12}$}}
\put(-20,95){\mbox{$k_{23}$}}

\put(-2,-10){\mbox{$0$}}  \put(8,-10){\mbox{$1$}}\put(18,-10){\mbox{$2$}}\put(28,-10){\mbox{$3$}}
\put(38,-10){\mbox{$4$}} \put(48,-10){\mbox{$5$}} \put(58,-10){\mbox{$6$}}
\put(26,-20){\mbox{$m$}}\put(54,-20){\mbox{$2m$}}

\put(-10,-3){\mbox{$0$}} \put(-10,7){\mbox{$1$}}\put(-10,17){\mbox{$2$}} \put(-10,27){\mbox{$3$}}
\put(-10,37){\mbox{$4$}}\put(-10,47){\mbox{$5$}}\put(-10,57){\mbox{$6$}}
\put(-25,27){\mbox{$m$}}\put(-30,57){\mbox{$2m$}}

\put(45,95){\mbox{$N=3,\ \   m=3$}}

\end{picture}
\caption{\footnotesize
The points, contributing to the sum (\ref{Psideco}) in the case of $sl_3$ ($N=3$) when the third root
is a sum of the two simple ones.
The sum in (\ref{Psideco}) is restricted to $1\leq k_{ij} \leq m$, which means that the coefficients
$k_{12}+k_{13}$ and $k_{23}+k_{13}$ in the linear combinations of $z_1-z_2$ and $z_2-z_3$ in the exponential
fill a peculiar hexagon on the plane $(k_{12},k_{23})$, where points in the interior are degenerate.
Degeneration degree  depends on the ``depth'' $d$, on the $d$-th hexagon $(k_{12}-d)(k_{23}-d)(m-k_{13}-d)=0$
all the points have degeneracy $d+1$.
The picture is drawn for $m=3$ and contains four embedded hexagons with $d=0,1,2,3=m$.
The dimension of the picture is equal to the number of simple roots, i.e. to $N-1=2$.
}
\label{hexas3}
\end{figure}

The sum \eqref{eq:n3psi-ansatz} actually goes over a peculiar hexagon on the $k_{12}-k_{23}$ plane, see Fig.\ref{hexas3},
where all internal points have multiplicities, equal to the depth of the corresponding sub-hexagons,
from $1$ to $m+1$. These degenerate coefficients, however, enter \eqref{eqs3} as
a monolithic unit, hence $\Psi$ is well-defined \textit{despite} the degeneracy.
The coefficients without multiplicities at the main contour of the hexagon are given by just
the products of $\psi$, thus one can substitute
\be \label{eq:psitil-ansatz}
\tilde\psi_{m|k_{12},k_{23},k_{13}} = 1 + k_{12}k_{23}(m-k_{13})\cdot \chi_{m|k_{12},k_{23},k_{13}}
\ee
so that the second term vanishes at the boundary, which
in the example Fig.~\ref{hexas3} is  marked by bigger black dots).

Indeed, on the  two straight sides of the enveloping hexagon near the origin we have $(k_{12},k_{23},k_{13})=(0,s,0)$
and  $(k_{12},k_{23},k_{13})=(0,s,0)$.
The label  $s$ runs from $0$ to $m$.
On the other two, parallel to them,
$(k_{12},k_{23},k_{13})=(2m,m+s,0)=\underline{(m,s,m)}$  and $(k_{12},k_{23},k_{13})=(m+s,2m,0)=\underline{(s,m,m)}$.
where only the last (underlined)representations belong to the cube  $0\leq k_{12},k_{23},k_{13}\leq m$.
Similarly the skew sides are parameterized by $(k_{12},k_{23},k_{13}) = (m+s,s,0)=\underline{(m,0,s)}$
and $(s,m+s,0)=\underline{(0,m,s)}$.
Thus in all these cases coefficient before $\chi_{m|k_{12},k_{23},k_{13}}$
in \eqref{eq:psitil-ansatz} vanishes and the term does not contribute.

\subsection{First sub-hexagon}

Now let us turn to the first {\it sub}-hexagon,
which in the case $m=3$ is denoted on Fig.~\ref{hexas3} by small white circles.

\subsubsection{$m=1$}

For $m=1$ it contains just a single doubly degenerate point, consisting of
two representatives in the cube:
$(k_{12},k_{23},k_{13}) = (1,1,0)$  and  $(k_{12},k_{23},k_{13}) =(0,0,1)$.
Equations (\ref{eqs3}), however, impose only one constraint for this point.
Thus we can put, say, one of the coefficients to zero and then obtain a clear expression for another one
\be
\chi_{1|0,0,1}=0, \nn \\
\chi_{1|1,1,0} = \frac{\L_1\L_2\L_3(q+1)^2(q-1)^3}{q(q\L_1-\L_2)(q\L_2-\L_3)(q\L_3-\L_1)}
\ee
However, the resulting combination $1+\chi_{1|1,1,0}$ does not look too nice .
But one can notice that it simplifies greatly and even factorizes
at $\L_2=q\L_3$ and/or at $\L_2=q^{-1}\L_3$,
which are zeroes of $\psi^{(1,2)}_{1,0}\psi^{(2,3)}_{1,0}\psi^{(1,3)}_{0,1}$.
This allows one to choose another gauge:
\be
\psi^{(1,2)}_{1|1}\psi^{(2,3)}_{1|1}\psi^{(1,3)}_{0|0}(1+\chi_{1|1,1,0})
+ \psi^{(1,2)}_{1|0}\psi^{(2,3)}_{1|0}\psi^{(1,3)}_{0|1}
= \psi^{(1,2)}_{1|1}\psi^{(2,3)}_{1|1}\psi^{(1,3)}_{0|0}\cdot \tilde \psi_{1|1,1,0}
+ \psi^{(1,2)}_{1|0}\psi^{(2,3)}_{1|0}\psi^{(1,3)}_{0|1}\cdot \tilde \psi_{1|0,0,1}
\ee
where
\be
\tilde \psi_{1|1,1,0} = \frac{(q^2\L_3-\L_1)(q\L_1-\L_3)}{q(q\L_3-\L_1)(\L_1-\L_3)}
\nn \\
\tilde \psi_{1|0,0,1} = \frac{(q^2\L_1-\L_3)(q\L_3-\L_1)}{q(\L_3-q\L_1)(\L_1-\L_3)}
\ee
Note that these $\tilde\psi$ are independent of $\L_2$, and depend only on the ``indices'' of non-simple root: $1$ and $3$.

\subsubsection{$m=2$}

At $m=2$ the first sub-hexagon contains six double points, and our first gauge corresponds to the choices
\be
\chi_{2|0,0,1}=0, & \chi_{2|1,1,0} = \frac{(q^2+q+1)^2(q-1)^3\L_1\L_2\L_3}
{2q(q^2\L_1-\L_2)(q^2\L_2-\L_3)(q\L_3-\L_1)}
\nn \\
\chi_{2|0,1,1} = 0, & \chi_{2|1,2,0} = \frac{(q^2+q+1)^2(q+1)(q-1)^3\L_1\L_2\L_3}
{4q^2(q^2\L_1-\L_2)(q\L_2-\L_3)(q\L_3-\L_1)}
\nn \\
\chi_{2|1,0,1} = 0, & \chi_{2|2,1,0} = -\frac{(q^2+q+1)^2(q+1)(q-1)^3\L_1\L_2\L_3}
{4q^2(q\L_1-\L_2)(q^2\L_2-\L_3)(q\L_3-\L_1)}
\ee
for the first three pairs (left lower side).
The other three pairs (at the right upper side) could include,
say  $\xi_{2|3,2,0}$, but this point lies beyond the cube $1\leq k_{ij}\leq m=2$,
therefore the proper choice is
\be
\chi_{2|1,0,2}=0, & \chi_{2|2,1,1} = \frac{(q^2+q+1)^2(q-1)^3\L_1\L_2\L_3}
{2q(q\L_1-\L_2)(q^2\L_2-\L_3)(q^2\L_3-\L_1)}
\nn \\
\chi_{2|0,1,2}=0, & \chi_{2|1,2,1} = \frac{(q^2+q+1)^2(q-1)^3\L_1\L_2\L_3}
{2q(q^2\L_1-\L_2)(q\L_2-\L_3)(q^2\L_3-\L_1)}
\nn \\
\chi_{2|1,1,2} = 0, & \chi_{2|2,2,1} = -\frac{(q^2+q+1)^2(q+1)(q-1)^3\L_1\L_2\L_3}
{4q^2(q\L_1-\L_2)(q\L_2-\L_3)(q^2\L_3-\L_1)}
\ee

\noindent The asymmetries between different lines
already start to creep in: the numeric prefactors are
the strange $\frac{1}{2}$, $\frac{1}{2}$, $-\frac{1}{4}$ and
$q$-powers in the Pochhammer-like factors $(q^\bullet \L_\bullet - \L_\bullet)$
differ.

This becomes even more prominent in the now present second sub-hexagon,
consisting of a single {\it  triple}
(therefore selecting coefficient \textit{a la} \eqref{eq:psitil-ansatz}
is not required)
for which we choose
\be
&\chi_{2|0,0,2} = \chi_{2|1,1,1} = 0, \nn \\
 &\chi_{2|2,2,0} =
\frac{(q^2+q+1)^2(q+1)^3(q-1)^4\L_1\L_2\L_3
\Big( \big((q^2+1)^3-2q^3\big)\L_1\L_2\L_3 - q^3\big(\L_1^2\L_2+\L_1\L_2^2
+ \L_1^2\L_3+\L_1\L_3^2 +\L_2^2\L_3+\L_2\L_3^2\big)
\Big)
}{8q^2(a\L_1-\L_2)(q^2\L_1-\L_2)(q\L_2-\L_3)(q^2\L_2-\L_3)
(q\L_3-\L_1)(q^2\L_3-\L_1)}
\label{chi2220}
\ee
Wherefrom it is clear that the formulas in the gauge for $\chi$ are indeed
do not reveal the nice structure.

However, the $\tilde\psi$-gauge provides factorized formulas and is quite nice.
We will write the expressions for the first sub-contour/shell for arbitrary $m$.

\subsubsection{First sub-hexagon, summary}

In general, this sub-hexagon consists of points $(a,b,c)$ with  $(a-1)(b-1)(m-1-c) =0$ and their companions $(a-1,b-1,c+1)$.
The values of the corresponding $\tilde \psi$ are:
\be
\boxed{
\tilde \psi_{m|a,b,c} = \frac{\big(q^{m-c}\L_1-\L_3\big)\big(\L_1-q^{a+b}\L_3\big)}
{q^{a+b-c-1}\big(q^{m+1-a-b}\L_1-\L_3\big)\big(\L_1-q^{c+1}\L_3\big)}
}
\ee
and
\be
\boxed{
\tilde \psi_{m|a-1,b-1,c+1} = \frac{\big(q^{2m+2-a-b}\L_1-\L_3\big)\big(\L_1-q^{c+1}\L_3\big)}
{q^{c+1}\big(q^{m-c}\L_1-\L_3\big)\big(q^{m+1-a-b}\L_1- \L_3\big)}
}
\ee

\subsection{Second sub-hexagon}

There are now three factorized items, associated with every dot.
This time only one of three $\tilde\psi$'s has pairs of elementary products in the numerator and denominator, while the other two contain products of four.

For $m=2$,
i.e. in the simplest case of (\ref{chi2220}), for the only triple point, we get:
\be \label{eq:psi3-cases-first}
\tilde \psi_{2|2,2,0} = \frac{(q\L_1-\L_3)(q^2\L_1-\L_3)(\L_1-q^3\L_3)(\L_1-q^4\L_3)}
{q^3(\L_1-\L_3)(\L_1-q\L_3)^2(\L_1-q^2\L_3) }
\nn \\
\tilde \psi_{2|1,1,1} = \frac{(q^2\L_1-\L_3)(\L_1-q^2\L_3)}{q(q\L_1-\L_3)(\L_1-q\L_3)}
\nn \\
\tilde \psi_{2|0,0,2} =  \frac{(q^3\L_1-\L_3)(q^4\L_1-\L_3)(\L_1-q \L_3)(\L_1-q^2\L_3)}
{q^3(\L_1-\L_3)(q\L_1- \L_3)^2(q^2\L_1- \L_3) }
\ee

This pattern continues for higher $m$. Indeed, for $m=3$ the second sub-hexagon
consists of four triples (denoted by small black dots in Fig.~\ref{hexas3}):
\be
\tilde \psi_{3|2,2,0} = \frac{(q^2\L_1-\L_3)(q^3\L_1-\L_3)(\L_1-q^3\L_3)(\L_1-q^4\L_3)}
{q^4(q\L_1-\L_3)(\L_1-\L_3)(\L_1-q\L_3)(\L_1-q^2\L_3) }
\nn \\
\tilde \psi_{3|1,1,1} = \frac{(q^4\L_1-\L_3)(\L_1-q^2\L_3)}{q^2(q^2\L_1-\L_3)(\L_1- \L_3)}
\nn \\
\tilde \psi_{3|0,0,2} =  \frac{(q^5\L_1-\L_3)(q^6\L_1-\L_3)(\L_1-q \L_3)(\L_1-q^2\L_3)}
{q^3(q^3\L_1-\L_3)(q^2\L_1- \L_3)^2(q\L_1- \L_3) }
\ee

\be
\tilde \psi_{3|2,3,0} = \tilde \psi_{3|3,2,0} = \frac{(q^2\L_1-\L_3)(q^3\L_1-\L_3)(\L_1-q^4\L_3)(\L_1-q^5\L_3)}
{q^5(\L_1-\L_3)(\L_1-q\L_3)^2 (\L_1-q^2\L_3) }
\nn \\
\tilde \psi_{3|1,2,1} = \tilde \psi_{3|2,1,1} = \frac{(q^3\L_1-\L_3)(\L_1-q^3\L_3)}{q^2(q\L_1-\L_3)(\L_1- q\L_3)}
\nn \\
\tilde \psi_{3|0,1,2} = \tilde \psi_{3|1,0,2} =
\frac{(q^4\L_1-\L_3)(q^5\L_1-\L_3)(\L_1-q \L_3)(\L_1-q^2\L_3)}
{q^3(q^3\L_1-\L_3)(q^2\L_1- \L_3)(\L_1-\L_3)(q\L_1- \L_3) }
\ee

\be
\tilde \psi_{3|2,3,1} = \tilde \psi_{3|3,2,1} = \frac{(q^2\L_1-\L_3)(q\L_1-\L_3)(\L_1-q^4\L_3)(\L_1-q^5\L_3)}
{q^3(\L_1-\L_3)(\L_1-q\L_3) (\L_1-q^2\L_3)(\L_1-q^3\L_3) }
\nn \\
\tilde \psi_{3|1,2,2} = \tilde \psi_{3|2,1,2} = \frac{(q^3\L_1-\L_3)(\L_1-q^3\L_3)}{q^2(q\L_1-\L_3)(\L_1- q\L_3)}
\nn \\
\tilde \psi_{3|0,1,3} = \tilde \psi_{3|1,0,3} =
\frac{(q^4\L_1-\L_3)(q^5\L_1-\L_3)(\L_1-q^2 \L_3)(\L_1-q^3\L_3)}
{q^5(q^2\L_1-\L_3)(q\L_1- \L_3)^2( \L_1-\L_3)  }
\ee

\be
\tilde \psi_{3|3,3,1}   = \frac{(q^2\L_1-\L_3)(q\L_1-\L_3)(\L_1-q^5\L_3)(\L_1-q^6\L_3)}
{q^3 (\L_1-q\L_3) (\L_1-q^2\L_3)^2(\L_1-q^3\L_3) }
\nn \\
\tilde \psi_{3|2,2,2}   = \frac{(q^2\L_1-\L_3)(\L_1-q^4\L_3)}{q^2(\L_1-\L_3)(\L_1- q^2\L_3)}
\nn \\
\tilde \psi_{3|1,1,3}   =
\frac{(q^3\L_1-\L_3)(q^4\L_1-\L_3)(\L_1-q^2 \L_3)(\L_1-q^3\L_3)}
{q^4(q^2\L_1-\L_3)(q\L_1- \L_3)(\L_1-\L_3)( \L_1-q\L_3)  } \label{eq:psi3-cases-last}
\ee

The formulas \eqref{eq:psi3-cases-first}-\eqref{eq:psi3-cases-last} are compatible with
\be
\tilde\psi_{m|a,b,c} =  \frac{(q^{m-c-1}\L_1-\L_3)(q^{m-c}\L_1-\L_3)(\L_1-q^{a+b-1}\L_3)(\L_1-q^{a+b}\L_3)}
{q^{2m-2c-1} (\L_1-q^{a+b+1-2m}\L_3) (\L_1-q^{a+b-4}\L_3)(\L_1-q^{c+1}\L_3)(\L_1-q^{c+2}\L_3) }
\nn \\
\tilde\psi_{m|a-1,b-1,c+1} = \frac{(q^{2m+2-a-b}\L_1-\L_3)(q^{2-a-b}\L_1-\L_3)}
{(q^{m+1-a-b}\L_1-\L_3)(q^{m+3-a-b}\L_1-\L_3)}
\nn \\
\tilde\psi_{m|a-2,b-2,c+2} = \frac{(q^{2m+4-a-b}\L_1-\L_3)(q^{2m+3-a-b}\L_1-\L_3)(\L_1-q^{c+2}\L_3)(\L_1-q^{c+1}\L_3)}
{q^{2m+2} (q^{2m-a-b}\L_1 - \L_3) (q^{5-a-b}\L_1- \L_3)(\L_1-q^{c-m}\L_3)(\L_1-q^{c+1-m}\L_3) }
\ee
at $(a-2)(b-2)(m-2-c) = 0$.

\def\lbb{\left[\!\left[ }
    \def\rbb{\right]\!\right] }

Continuing along these lines, one arrives at the formula (36) in \cite{MMP1}
for \textit{all} shells:
\begin{align} \label{eq:psi-tilde-general}
  \widetilde{\psi}_{m|a,b,c} =
  \frac{\lbb2 m-a-b\rbb}{\lbb2 m -a-b-c\rbb}
  \cdot
  \frac{\lbb m-1-a-b-c\rbb}{\lbb m -1-a-b\rbb}
  \cdot
  \frac{\lbb m-a-b\rbb}{\lbb -1-a-b\rbb}
  \cdot
  \frac{\lbb -1-a\rbb}{\lbb m-a\rbb}
  \cdot
  \frac{\lbb -1-b\rbb}{\lbb m-b\rbb}
  \cdot
  \frac{\lbb m-c\rbb}{\lbb m-1\rbb},
\end{align}
where $\lbb n \rbb$ is the infinite downward Pochhammer-like product
\begin{align}
  \lbb n \rbb = \prod_{i=0}^\infty \left(q^{n - i} \L_1 - \L_3\right)
\end{align}
Quite surprisingly, the shell structure ultimately turns out
to be less important than could be anticipated from the outset. The reasons
and implications of this are yet to be fully understood and appreciated.

\section{Comparison with Rodrigues type formula
\label{genarg}}

One of possible representations for solutions to the  linear conditions  (\ref{linsys0})
\be
\Psi_m(z_k+j,\vec\lambda) =\Psi_m(z_l+j,\vec\lambda), \ \ \ 1\leq j \leq m
\ee
is \cite[Eq.(3.16)]{Cha}
\be\label{prod}
\Psi_m(z_k,\vec\lambda)=\prod_{\substack{k_{ij}=0 \\ \sum_{i<j}k_{ij}^2\ne 0}}^m\left(\hat D_\Z-q^{-{m\over 2}}\sum q^{\lambda_i+\sum_{j>i}k_{ij}-\sum_{j<i}k_{ji}}\right)
\left[q^{(\vec\lambda+m\vec\rho)\cdot\vec z}\Delta^{-1}_{q,t}(\Z)\right]
\ee
where the product runs over all \textit{non-zero} multiplicity vectors for positive roots
$k_{ij}$, $i<j, i,j=1..N$; similarly to \eqref{Psideco}, where the sum
is taken over a similar set but \textit{including} the point where {\it all} $k_{ij}=0$.

The difference operator $\hat{D}_\Z$
\be
\hat D_\Z:=\sum_{i=1}^N\prod_{j\ne i}{\Z_j-q^{-m}\Z_i\over \Z_j-\Z_i}e^{\p_{z_i}}
\ee
is nothing but the Ruijsenaars-Macdonald operator, acting here on the familiar inverse
\be
\Delta_{q,t}^{-1}(\Z):=\prod_{i\ne j}\prod_{k=1}^m\left(q^k{\Z_i\over \Z_j}-1\right)
\ee
of the $q,t$-deformed square of the Vandermonde determinant at $t=q^{-m}$ \cite{Mac}. The normalization of (\ref{prod}) is such that the coefficient in front of $q^{(\vec\lambda+m\vec\rho)\cdot\vec z}$ turns out to be
\be
q^{m\over 2}\prod_{k>l}\prod_{j=1}^m\Big(q^{\lambda_k-\lambda_l-j}-1\Big)\Big(1-q^{j}\Big)
\ee
which means that it differs from the symmetric normalization by monomial factor
$
\prod_{k>l}q^{-{m(\lambda_k-\lambda_l)\over 2}}
$.
However, the expression (\ref{prod}) is rather involved, and does not allow one to read off the coefficients of the BAF immediately.

This is because \eqref{Psideco} and \eqref{prod} are structurally very different:
the former features the \textit{sum} over space of positive root multiplicities,
while the latter features the \textit{product}. If one imagines representing square-bracketed
part of \eqref{prod} as explicit (extra) sum over $\Z$-monomials and, further, expands
product of $(\hat{D}_\Z - ...)$ factors in Newton binomial-like fashion (again, involving
more extra summations) then one summand in \eqref{Psideco} is assembled out of
many contributions from \eqref{prod}. Moreover, just as \eqref{Psideco},
the formula \eqref{prod} has the same structural problem: $\hat{D}_\Z$-operators
do not distinguish identical $\Z$-monomials, occurring from different (linearly dependent)
combinations of positive roots.
Hence, the problem of clever choice of individual $\tilde{\psi}$ coefficients
(and understanding its meaning and significance) remains relevant.

\section{Solving defining linear equations for twisted BAF
\label{twBAF1}}

In the Appendix to \cite{ChE} O. Chalykh proposed a generalization of the BAF's based on their relevance to bilinear integral formulas: these functions appear naturally when one studies more general coefficient in the Gaussian integral weight.
Strikingly, these functions satisfy an ansatz + a set of difference equations very similar to those for the BAF's.

That is, for $N=2$ these twisted BAF's with the slightly changed normalization so that they depend only on the ratio $x=x_1/x_2$,
\be \label{eq:twba-ansatz}
\overline{\Psi}^{(a)}_m(\Z) := x^{\lambda\over a}\Z^{\frac{ ma}{2}}\sum_{k=0}^{am} \Z^{-k} \psi^{(a)}_{m,k}
\ee
satisfy the system of $ma$ linear equations
\be
 \overline{\Psi}^{(a)}_m \left(e^{-\frac{2\pi i}{a}s} q^{\frac{j}{a}}\right) =
e^{\frac{2\pi i}{a}sj}\cdot \overline{\Psi}^{(a)}_m \left(e^{-\frac{2\pi i}{a}s} q^{-\frac{j}{a}}\right)
\label{linsys}
\ee
for $s=1,\ldots,a$ and $j=1,\ldots m$.

Solutions for small $a$ are given by \cite{MMP1}
\be
\psi^{(1)}_{m,k} = (-)^k q^{-mk+\frac{k(k-1)}{2}} \frac{[m]!}{[m-k]![k]!}
\prod_{i=1}^k \frac{[\lambda+m-i+1]}{[\lambda-i]}
\label{psi1ans}
\ee
and
\be
\psi^{(2)}_{m,k} = \sum_{r={\rm max}(0,k-m)}^{{\rm floor}\left(\frac{k}{2}\right)}
(-)^{k+r}q^{(\frac{k}{2}-r)(\lambda-2m-1)-mr+\frac{r(r-1)}{2}}
\frac{[m+k-2r]!}{[m-k+r]![k-2r]![r]!} \frac{\prod_{i=1}^r [\lambda+m-i+1]}{\prod_{i=1}^{k-r} [\lambda-i]}
\label{psi2ans}
\ee
with $[k]:=\frac{q^k-1}{q-1}$ and $[n]! = \prod_{k=1}^n [k]$.
If $[n]!$ for negative $n$ is defined to have poles, then the sum over $r$ is automatically restricted to $r\geq k-m$,
but we wrote this explicitly to avoid possible confusion.

\bigskip

For arbitrary $a$
\be
\psi^{(a)}_{m,k} =  q^{ \lambda\, \left\langle1-\frac{k}{a}\right\rangle}\cdot q^{-\frac{(a-1)k^2 }{a}}\cdot
\sum_{r=0}^m \frac{c^{(a)}_{m,k,r}}{\prod_{i=1}^{r}[\lambda-i]},
\label{psivsc}
\ee
where $\left\langle x \right\rangle$ denotes fractional part of $x$.
This means that the formula exhibits ``jumps'' at division lines $k = n a$,
and in the region $a \geq k$ there are no such jumps. Such behavior is typical for
root-of-unity equations like \eqref{linsys}, and it is no surprise we observe it here.

\bigskip

Manifestly, one has
\be
c^{(3)}_{1,k,r} = \left[  \begin{array}{l|cc}
k\backslash r & 0 & 1 \\
\hline  && \\
1 &  0 & -\frac{[2]}{q} \\ && \\
2 & 0 & -q[2]\\  && \\
3 & -q^7 & -q^5[2] \\
\end{array} \right]
\ \ \ \
c^{(4)}_{1,k,r} = \left[  \begin{array}{l|cc}
k\backslash r & 0 & 1 \\
\hline  && \\
1 &  0 & -\frac{[2]}{q} \\ && \\
2 & 0 & -q[2]\\  && \\
3 & 0 & -q^5[2] \\ && \\
4 & -q^{13} & -q^{11}[2] \\
\end{array} \right]
\ \ \ \
c^{(5)}_{1,k,r} = \left[  \begin{array}{l|cc}
k\backslash r & 0 & 1 \\
\hline  && \\
1 &  0 & -\frac{[2]}{q} \\ && \\
2 & 0 & -q[2]\\  && \\
3 & 0 & -q^5[2]\\  && \\
4 & 0 & -q^{11}[2]\\  && \\
5 & -q^{21} & -q^{19}{[2]} \\
\end{array} \right]
\ee
i.e.
\be
\boxed{
c^{(a)}_{1,k,r} = \left\{\begin{array}{c||cc|c}
r && 0 & 1  \\ \hline &&&\\
1\leq k <a  &&
0  &   - q^{k^2-k-1}[2] \\ &&& \\
k=a && -  q^{a^2-a+1}    & - q^{a^2-a-1}[2]
\end{array} \right.
}
\ee
or
\be
c^{(a)}_{1,k,r} =  -  q^{a^2-a+1}\delta_{r,0}  \delta_{k,a} - q^{k^2-k-1}[2] \delta_{r,1}
= -q^{k^2-k}\left(q \,\delta_{r,0}  \delta_{k,a} + \frac{[2]}{q} \delta_{r,1}\right)
\nn \\ \label{eq:ca-1}
\ee

{\footnotesize
\be
\!\!\!\!\!\!\!\!\!\!\!\!\!\!\!\!\!\!\!\!\!\!\!\!\!\!\!\!\!\!\!\!\!\!\!\!\!\!\!\!\!\!
  c^{(3)}_{2,k,r} =    \left[\begin{array}{l|ccc}
k\backslash r & 0 & 1 & 2 \\
\hline  &&&\\
1 & 0  & -\frac{[2][3]}{q^2} & 0 \\ &&&\\
2 & 0 & \frac{[2][3](q^3-q^2-1)}{q}& \frac{[3][4]}{q}\\ &&&\\
3 & -q^7[2] &  q^2[2][3](q^4-q^2-1) &  q^2[2][3][4]  \\ &&&\\
4 &q^{11} [2][3](q-1) &\!\! q^7[2][3](q^5+q^4+q^3-q^2-1)\!\!\!\!\!  & q^7[3]^2[4]  \\ &&&\\
5 &q^{17} [2][3](q-1)& q^{14}[2][3](q^4+q^3-1)&q^{14}[2][3][4]  \\ &&&\\
6 &q^{27} &q^{24}[2][3]& q^{21} [3][4]  \\
\end{array}\right]
\ \
  c^{(4)}_{2,k,r} =    \left[\begin{array}{l|ccc}
k\backslash r & 0 & 1 & 2 \\
\hline  &&&\\
1 & 0  & -\frac{[2][3]}{q^2} & 0 \\ &&&\\
2 & 0 & \frac{[2][3](q^3-q^2-1)}{q}& \frac{[3][4]}{q}\\ &&&\\
3 &0 &q^2[2][3](q^4-q^2-1) & q^2[2][3][4]  \\ &&&\\
4 &-q^{13}[2] & q^7[2][3](q^5-q^2-1)  & q^7[3]^2[4]  \\ &&&\\
5 &q^{19}[2][3](q-1) & \!\!q^{14}[2][3](q^6+q^5+q^4-q^2-1)\!\! &q^{14}[3][4]^2 \\ &&&\\
6 &q^{27}[2][3](q-1)& \!\! q^{23}[2][3](q^5+q^4+q^3-q^2-1)\!\! &q^{23}[3]^2[4]  \\ &&&\\
7 &q^{37}[2][3](q-1)& q^{34}[2][3](q^4+q^3-1)&q^{34}[2][3][4] \\ &&&\\
8 &q^{51} &q^{48}[2][3]& q^{45}[3][4] \\
\end{array}\right]
\nn
\ee
}

{\footnotesize
\be
\!\!\!\!\!\!\!\!\!\!\!\!\!\!\!\!\!\!\!\!\!\!\!\!\!\!\!\!\!\!\!\!\!\!\!\!
  c^{(5)}_{2,k,r} =    \left[\begin{array}{l|ccc}
k\backslash r & 0 & 1 & 2 \\
\hline  &&&\\
1 & 0  & -\frac{[2][3]}{q^2} & 0 \\ &&&\\
2 & 0 & \frac{[2][3](q^3-q^2-1)}{q}& \frac{[3][4]}{q}\\ &&&\\
3 &0 &q^2[2][3](q^4-q^2-1) & q^2[2][3][4]  \\ &&&\\
4 &0 & q^7[2][3](q^5-q^2-1)  & q^7[3]^2[4]  \\ &&&\\
5 &-q^{21}[2]& q^{14}[2][3](q^6 -q^2 -1)&q^{14}[3][4]^2 \\ &&&\\
6 &q^{29}[2][3](q-1) & q^{23}[2][3](q^7+q^6+q^5-q^2-1)&q^{23}[3] [4][5]  \\ &&&\\
7 &q^{39}[2][3](q-1)& q^{34}[2][3](q^6+q^5+q^4-q^2-1)&q^{34}[3][4]^2 \\ &&&\\
8 &q^{51}[2][3](q-1)& q^{47}[2][3](q^5+q^4+q^3-q^2-1)&q^{47}[3]^2[4]  \\ &&&\\
9 &q^{65}[2][3](q-1)& q^{62}[2][3](q^4+q^3-1) &q^{62}[2][3][4] \\ &&&\\
10 &q^{83} &q^{80}[2][3]& q^{77}[3][4] \\
\end{array}\right]
\nn
\ee
}

\bigskip

i.e.

{\footnotesize
\be
\boxed{
c^{(a)}_{2,k,r} =  \left\{\begin{array}{c||cc|c|c}
r && 0 & 1 & 2 \\ \hline &&&& \\
1\leq k \leq a &&
 - q^{a^2-a+1} [2]\delta_{k,a}
  & q^{k^2-2k-1}  [3]\Big(q^{k+1}[2]-[4]\Big)
 &  q^{k^2-2k-1} [3][4][k-1] \\ &&&& \\
  a<k< 2a &&
 q^{k^2-3k+2a+1}[2](q^3-1)
 &  q^{k^2-2k-1}  [3]\Big(q^{2a+1-k}[2][3]-[4]\Big)
 &   q^{k^2-2k-1}[3][4][2a+1-k]
 \\ &&&& \\
k=2a &&  q^{4a^2-4a+3} &  q^{4a^2-4a } [2][3]    & q^{4a^2-4a-3}[3][4]
\end{array} \right.
}
\nn
\ee
}

\bigskip

or

\bigskip

{\footnotesize
\be
\boxed{
c^{(a)}_{2,k,r} =  q^{k^2-2k}\cdot \left\{\begin{array}{c||cc|c|c}
r && 0 & 1 & 2 \\ \hline &&&& \\
1\leq k \leq a &&
  -q^{ a+1} [2]\delta_{k,a}
  & q^{ -1}  [3]\Big(q^{k+1}[2]-[4]\Big)
 &  q^{ -1} [3][4][k-1] \\ &&&& \\
  a<k< 2a &&
 q^{2a+1-k}[2](q^3-1)
 &  q^{ -1}  [3]\Big(q^{2a+1-k}[2][3]-[4]\Big)
 &   q^{ -1}[3][4][2a+1-k]
 \\ &&&& \\
k=2a &&  q^{ 3} &   [2][3]    & q^{ -3}[3][4]
\end{array} \right.
}
\label{eq:c2-tab}
\ee
}

\bigskip

Next, for $m=3$

\bigskip

{\footnotesize
\be
c^{(a)}_{3,k,r} =   -q^{k^2-3k}\cdot
\nn
\ee
\be
\!\!\!\!\!\!\!\!\!\!\!\!\!\!\!\!\!\!\!\!\!\!\!\!\!\!\!\!\!
\boxed{
 \left\{\begin{array}{c||c|c|c|c}
r & 0 & 1 & 2 & 3 \\ \hline &&&& \\
1\leq k \leq a &
  q^{ 2a+1} [3]\delta_{k,a}
  & \left(q^{2k+2}-q^k[5]+\frac{[5][6]}{q[2][3]}\right)[3][4]
  & \frac{[4][5][k-1]}{q}\Big(q^{k+1}[3]-[6]\Big)
 & \frac{[4][5][6][k-1][k-2]}{[2]}  \\ &&&& \\
  a<k< 2a &
 -q^{2a+2} (q^3-1)[4]\cdot
 &  -\left(q^k -\frac{[6]}{q[2][3]}+q^{3a-k}[4]\right)[5] +
 & -\Big(q^{2a+1}[k-2a-1]+\frac{q^3+1}{q}[k-1] +
 &  \Big(-q^{k-2}[2a+2-k][a+1-k] + \!\!\!\!
 \\ &\cdot \Big(q^{a-k-1}[5]-[3]\Big) & +q^{2a}\frac{[3]^2[4]}{[2]}
 &+q^{2a-1}[5][a+1-k]\Big)  [3][4][5] &+\frac{[a][a-1]}{[2]}\Big) [4][5][6]\\
  &&&& \\
k=2a &  -q^{2a+3}[3] &   \left(q^{2a+1}[3]-q^a[5]\right)[3][4]
& \frac{q^{a+3}[2][a-1]-q^2[a-1]-1}{q^3}[3][4][5] & \frac{q^{a-2}[3][2]+[a-2]}{q^3[2]}[4][5][6][a-1]=
\\
&&&& =\left(\frac{[a+1][a+2]}{q^3[2]} - q^{2a-5}[3]\right)[4][5][6]
\\ &&&&\\
 2a<k< 3a &
 q^{6a+3-2k}(q^3-1)[4] & \left(-q^{3a}[5]+q^{6a+1-k}\frac{[3][4]}{[2]}\right)q^{-k}[3][4]
 &   \left([3][4]q^{3a+2-k}-[6]\right)\frac{[4][5][3a+1-k]}{q^3[2]}
 & \frac{[4][5][6][3a+2-k][3a+1-k]}{q^3[2]}
 \\ &&&& \\
k=3a &  q^{ 6} &   q^2[3][4]    & \frac{[3][4][5]}{q^2} &  \frac{[4][5][6]}{q^6}
\end{array} \right.
}
\nn
\ee
}

\bigskip

\noindent
Generalization to higher $m$ is straightforward, but the answers are too big to be presented here.
Some structures are already seen even in these simplest examples,
and are partially elaborated on in the next section.
Still, it is an open question how to use them for converting the result into a reasonably compact form.

\section{Comment on scalable way to program linear equations for twisted BAF's
\label{twBAF2}}

\newcommand\lla{\left\langle}
\newcommand\rra{\right\rangle}

The $a$-th roots of unity, appearing in many places in equation \eqref{linsys}
raise the questions about suitable branch/sign choices. Moreover, available CAS software
(MAPLE, Mathematica etc.) does not always work well with such roots, and simplifies them wrongly/takes too long to do so. The goal of this section is to demonstrate that
defining equations for twisted BAF can be worked with purely \textit{symbolically},
and in logarithmic ($\Z_i = q^{z_i}, \ \ \L_i = q^{\lambda_i}$) terms;
without the need to crunch through $\exp{(i/a k)}$ factors manually.

\bigskip

The Chalykh original definition for the twisted BAF ansatz reads
\begin{align}
  \Psi^{(a)}(z,\lambda)= q^{\lla\lambda,z\rra/a}
  \sum_{\nu \in \mathcal{N} \cup \rho + a^{-1} P}
  \psi_\nu(\lambda) q^{\lla\nu,z\rra}
  ,
\end{align}
where, \textit{crucially} $\mathcal{N}$ is a polygon over
the \textit{real} numbers
(which is not too obvious from Chalykh's notation),
which gets intersected by
the lattice of \textit{rational} points, whose coordinates
are in steps of $1/a$.

\bigskip

For $N=2$ this ansatz manifestly gives (to be compared with \eqref{Psideco},
where the gauge $\Z = q^{z_1 - z_2}$ was chosen)
\begin{align}
  \Psi^{(a),[2]}_m(z_1,z_2) =
  q^{\frac{\lambda_1 z_1}{a} + \frac{\lambda_2 z_2}{a}}
  \sum_{\nu=0}^{m a} \psi_{m,\nu}(\lambda) q^{-m/2 (z_1-z_2) + \frac{\nu}{a}(z_1-z_2)}
\end{align}
Difference equations read
\begin{align}
  \Psi^{(a)}(z - \frac{1}{2}j\alpha) = \epsilon^j \Psi^{(a)}(z + \frac{1}{2}j\alpha),
  \ \ \ \text{ with } q^{{\lla\alpha,z\rra\over 2}} = \epsilon,
  \text{ and } \epsilon^a = 1,
\end{align}
where $z = (z_1,z_2)$ is a two-dimensional vector; and for the only root
$\alpha = e_1 - e_2$ in case of $N=2$ we have
\begin{align}
  z \pm \frac{1}{2}j\alpha
  = \left(\begin{array}{c}z_1\pm\frac{1}{2}j \\ z_2\mp\frac{1}{2}j \end{array}\right),
  \text{ with }  q^{\frac{1}{a}(z_1 - z_2)} = \epsilon
\end{align}
therefore equations are
\begin{align}
  \underbrace{q^{\frac{\lambda_1 z_1}{a} + \frac{\lambda_2 z_2}{a}}}_{\text{can cancel}}
  q^{\frac{(- j) (\lambda_1 - \lambda_2)}{2 a}}
  \sum_{\nu=0}^{m a} \psi_\mu(\lambda) q^{mj/2 - \nu j/a} \epsilon^\nu
  = \epsilon^j
  \underbrace{q^{\frac{\lambda_1 z_1}{a} + \frac{\lambda_2 z_2}{a}}}_{\text{can cancel}}
  q^{\frac{j (\lambda_1 - \lambda_2)}{2 a}}
  \sum_{\nu=0}^{m a} \psi_\mu(\lambda) q^{-mj/2 + \nu j/a} \epsilon^\nu
\end{align}
Since $\epsilon^{k+a} = \epsilon^k$, one needs to combine coefficients in front of
$\epsilon$-monomials whose exponents differ by $a$, obtaining as equations
the combined coefficients in front of independent monomials
$\epsilon^0, \cdots \epsilon^{a-1}$. Furthermore, log-variable $\L_i = q^\lambda_i$ may be introduced throughout.

This technique becomes especially important when analyzing more complicated root systems,
where roots of unity for different roots start interfere with one another.

\subsection{$m=1$}

Even though it appears that $\psi_\nu(\lambda)$ should depend on fractional
powers of $q$: $q^{1/a}$ and fractional exponentials of $\lambda_i$,
$q^{\lambda_i/a}$, in fact almost all fractionality cancels, and one gets

\begin{align}
  \psi^{(a)}_{1,k} = (-1) q \delta_{a,k}
  + \frac{q^{-k+k^2/a}(q^2-1) \left(\frac{\L_1}{\L_2}\right)^{k/a}}
  {\left(q \frac{\L_1}{\L_2} - 1\right)}
\end{align}
where $\L_i = q^{\lambda_i}, k = 1..a$, i.e. for every \ $\Psi^{(a)}$ we have only one
occurrence of simple contact term at $k=a$, in agreement with \eqref{eq:ca-1}.

\subsection{$m=2$}
In practice, to obtain such answer one first looks at ``stable''
answers in the region $a > k$, and then corrects them, gradually
increasing $\Delta=a-k$.

\begin{align}
  \psi^{(a)}_{2,k} = & \
  \underbrace{
  \frac{(-1)q^{k(a-k)}}{\left(\L_1^a - q^a \L_2^a\right)(q^{2 a}-1)
    \L_1^{a-k}\L_2^k}
  \cdot
  \frac{\left(q^{(k+1)a}
    \frac{(\L_1^{2a}-\L_2^{2a})}{(\L_1^{a}-\L_2^{a})}
    - \frac{(q^{4 a} -1)}{(q^{2 a} - 1)} \L_1^a
    \right)}{(q^a - 1)(-1)q^{k a}\left(\L_1^a - q^{2 a} \L_2^a\right)
  }}_{\text{regular part}} \\ \notag
  + & \sum_{\Delta=0}^{a-k} f(\Delta,a,k)
\end{align}
\begin{align}
  f(\Delta,a,k) = & \ \delta_{\Delta > 0}
  \underbrace{
    \frac{\left(q^{2 \Delta a} \L_2^{a+1}
      - q^{(2 \Delta-1)(a-1)} \L_1 \L_2^a
      + q^{a-(2 \Delta-1)} \L_1^{a+1}
      - \L_1^a \L_2 \right)
      (\L_1^{2 a} - \L_2^{2 a})
    }{(\L_1^{a} - \L_2^{a})(\L_1^{2 a} - q^a \L_2^{2 a})
      (\L_1^{2 a} - q^{2 a} \L_2^{2 a})}}_{\text{semi-regular part}}
  \\ \notag
  & \ \ \ \ \ \ \ \underbrace{\cdot \frac{(q^{3 a}-1)(q^{2 a} - 1)}{(q^a-1)}
    \frac{\L_2^{\Delta-1}}{\L_1^\Delta}
    \frac{(-1)}{q^{(\Delta-1)a - \Delta^2}}
  }_{\text{semi-regular part (continued)}} \\ \notag
  + & \underbrace{\delta_{\Delta,a} q^{a(m+1)}
  }_{\text{semi-regular contact term}} \\ \notag
  + & \ \delta_{\Delta,1}
  \frac{(\L_1^a - q^{- a} \L_2^a)
    (\L_1^a - q^{-2 a} \L_2^a)}
       {(\L_1^a - q^{a} \L_2^a)(\L_1^a - q^{2 a} \L_2^a)}
       \frac{(q^{2 a} - 1)}{(q^a-1) q^{4 a}}
   \\ \notag
  + & \ \delta_{a-k,0} (-1) q^a \frac{(q^{2 a}-1)}{(q^a-1)},
\end{align}
again, in agreement with corresponding table \eqref{eq:c2-tab}.
Here we see that the table \eqref{eq:c2-tab}, in fact, has more structure to it than
seems at first glance: the bulk of its content may be described by just two functions,
nicknamed here ``regular part'' and ``semi-regular part'' (i.e. the regular part of the
corrections), and $2 \mathop{=}^? m$ truly contact terms.

\subsection{Arbitrary $m$}

It is clear that complexity increases rapidly as one goes along $m=1,2,\dots$.
Therefore, one needs proper concise terms (\textit{aka} ``basis functions'')
in which to describe the answers.

Here we report a partial success in this direction, in the most simple and regular region
of the parameter space $a > k \geq m$.

This \textit{regular} answer is cleanly expressed in terms of basis functions
\begin{align}
  f_l = \frac{\L^{k/a} q^{- l k + k^2/a}}{\prod_{i=1}^l (q^i \L - 1)},
  \L = \frac{\L_1}{\L_2} \\ \notag
  \psi^{(a)}_{m,k,\text{reg}} = \sum_{j=1}^m f_j c_j(m,k),
\end{align}
where coefficients $c_j$ depend on $k$ only, not on $a$ and not on $\L$.

Explicitly

\begin{align}
  \psi^{(a)}_{1,k,\text{reg}} = & \ f_1 \cdot [2]!(q-1) \\ \notag
  \psi^{(a)}_{2,k,\text{reg}} = & \ f_1 \cdot [3]! \frac{(q-1)}{q}
  + f_2 \cdot \frac{[4]!}{[2]!} \frac{(q-1)}{q} \br{q^k - q} \\ \notag
  \psi^{(a)}_{3,k,\text{reg}} = & \ f_1 \cdot \frac{[4]!}{[2]!}
  \frac{(q-1)}{q^2}
  +
  f_2 \cdot \frac{[5]!}{[2]!} \frac{(q-1)}{q^3} \br{q^k - q}
  +
  f_3 \cdot \frac{[6]!}{[3]![2]!} \frac{(q-1)}{q^3} \br{q^k - q}\br{q^k - q^2}
  \\ \notag
  \psi^{(a)}_{4,k,\text{reg}} = & \ f_1 \cdot \frac{[5]!}{[3]!}
  \frac{(q-1)}{q^3}
  + f_2 \cdot \frac{[6]!}{[2]![2]!} \frac{(q-1)}{q^5} \br{q^k - q}
  + f_3 \cdot \frac{[7]!}{[3]![2]!} \frac{(q-1)}{q^6} \br{q^k - q}\br{q^k - q^2} + \\
  & + f_4 \cdot \frac{[8]!}{[4]![3]!} \frac{(q-1)}{q^6} \br{q^k - q}\br{q^k - q^2}\br{q^k - q^3}
\end{align}

What, at the moment, is already clear is that the overoptimistic hope one could have had
from successes \eqref{psiAB},\eqref{eq:psi-tilde-general}
for the non-twisted functions that one can choose fully
factorized coefficients $\psi$ is \textit{lost} in the twisted case. Instead, an extra
structure, summation $\sum$ of some fully factorized \textit{auxiliary} expressions
manifests itself. This is yet to be matched with explicit description of the
corresponding DIM Hamiltonians $\hat{H}^{(-1,a)}_k$ via Cherednik operators
\cite[sec.6.2]{MMP1}, which, on the surface, do not feature similar extra summation.

\section{Twisted BAF as eigenfunctions of the integer-ray Hamiltonians
\label{Hams}}

\subsection{Untwisted BAF and the peculiarities of the Hamiltonian eigenfunctions}

The BAF at $a=1$,
\be
\bar\Psi^{(1)}_m(\Z) = \Z^{\frac{m}{2}} \left(1+ \sum_{k=1}^m \Z^{-k}\psi^{(1)}_{m,k}\right)
\label{Psi1}
\ee
satisfies a system of $m$ equations
\be
Q^{\frac{j}{2}} \bar\Psi^{(1)}_m( q^j) =  Q^{-\frac{j}{2}} \bar\Psi^{(1)}_m(  q^{-j}), \ \ \ \ j=1,\ldots, m
\label{linsysPsi1}
\ee
with the solution (\ref{psi1ans}).
Note that particular {\it equations} in this Chalykh's linear system
(\ref{linsysPsi1}) are independent of $m$,
what depends is the {\it number} of equations in the system, and in result
the parameter $m$ distinguishes between its different solutions, i.e. order of the polynomial in (\ref{Psi1}).
Note a slight change of notation:
throughout this section, we extract the overall factor $\Z^{\frac{\lambda}{2}}$ from $\Psi$ in (\ref{Psi1})
and put explicit factors with $Q=q^{\frac{\lambda}{a}}$ into the equations (\ref{linsysPsi1}) instead.

At the same time the BAF appeared from Macdonald polynomial $M_R[x]$ with $[R|>2m$,
which were a set of common eigenfunctions for the Ruijsenaars Hamiltonians.
This means that besides (\ref{linsysPsi1}) it should satisfy another kind of equations.
In particular from the first Hamiltonian,
also first-order in the difference operator,
\be
\hat {\overline {H}}^{(0,1)}_1(z) = \frac{q^z t-1}{q^z-1} e^{\p_z}
+   \frac{q^z -t}{q^z-1}  e^{-\p_z}
\label{HRui1}
\ee
with the eigenvalue $q^{-{\lambda\over 2}}t^{\frac{1}{2}} (q^{\lambda}+1)$
which acts on $\bar\Psi_m(z):=Q^{\frac{z}{2}}\bar\Psi^{(1)}_m(\Z)$
we get
\be
 \hat {\overline {H}}^{(0,1)}_1(z)\bar\Psi_m(z) =
  \left({1-q^{z-m}\over 1-q^{z}}e^{\partial_z}+{1-q^{z+m}\over 1-q^{z}}q^{-m}e^{-\partial_z}\right)
\bar\Psi_m(z) =
q^{-{\lambda+m\over 2}}(q^{\lambda}+1) \overline{\Psi}_m(z,\lambda)
\label{Hrui1a}
\ee
or
\be
 \frac{\Z q^{-m}-1}{\Z-1}\,Q^{\frac{1}{2}}\bar\Psi^{(1)}_m(q\Z)
+   \frac{\Z q^{m}-1}{\Z-1}q^{-m}Q^{-\frac{1}{2}} \bar\Psi^{(1)}_m(q^{-1}\Z)
= q^{-\frac{m}{2}} ( Q^{\frac{1}{2}} + Q^{-\frac{1}{2}})\bar\Psi^{(1)}_m(\Z)
\label{1RPsi1}
\ee
where $\Z=q^z$ and $Q=q^\lambda$.
Note that the operators (\ref{Hrui1a}) and (\ref{1RPsi1}) are $m$-dependent,
a possible interpretation is that Macdonald {\it parameter} $t$ in (\ref{HRui1}) is now treated as
{\it an operator} $\hat t$, for which $\bar\Psi_m$ are peculiar eigenfunctions:
\be
\hat t \bar\Psi_m = q^{-m} \bar\Psi_m
\ee
Of course, apart from ``discrete spectrum'' solutions, provided by Macdonald polynomials,
the Ruijsenaars Hamiltonian(s) with $c$-number $t$ have a full set of (common) eigenfunctions
for continuous spectrum, but these, in contrast to the Macdonald polynomials are not
so well publicized; and are certain functions of (basic-) hypergeometric type:
the so-called Noumi-Shiraishi functions \cite{NS}.
The treatment of these continuous spectrum functions, which have their own set
of peculiarities is postponed to a separate publication \cite{MMP3}.

\subsection{On the relation between (\ref{linsysPsi1}) and (\ref{1RPsi1})}

The two systems of the first order difference equations (\ref{linsysPsi1}) and (\ref{1RPsi1})
look somewhat different.
Moreover, they can seem {\it very} different, because (\ref{1RPsi1}) contains  arbitrary $\Z$,
while in (\ref{linsysPsi1}) $\Z$ is constrained to just $m$ points  $\Z=q^j$.
Still the interplay between them can be made quite explicit, when the solution
is a polynomial in $\Z^{-1}$, with just $m$ coefficients to be fixed.

Namely, let us take $m=1$.
Then (\ref{linsysPsi1}) tells that $Q\bar\Psi^{(1)}_1(q)=\bar\Psi^{(1)}_1(q^{-1})$.
In the ansatz $\bar\Psi^{(1)}_1(\Z) = \Z^{1/2}+u\Z^{-1/2}$ we get $u=-\frac{Qq-1}{Q-1}$.
At the same time to determine a single parameter $u$ from (\ref{1RPsi1}) it is enough to consider it
at a single point.
Most convenient is the choice $\Z=1$, since then the r.h.s. does not contribute, while in the numerator at the l.h.s. we get
$
(q^{-1}-1) Q^{1/2} \bar\Psi^{(1)}_1(q) + (q-1)q^{-1}Q^{-1/2}\bar\Psi^{(1)}_1(q^{-1})=0,
$
i.e. \textit{exactly the same} equation (\ref{linsysPsi1}).

For $m=2$ analysis is a little trickier.
Now (\ref{linsysPsi1}) consists of two equations: the same
$Q\bar\Psi^{(1)}_2(q)=\bar\Psi^{(1)}_2(q^{-1})$ and $Q^2\bar\Psi^{(1)}_2(q^2)=\bar\Psi^{(1)}_2(q^{-2})$.
They are sufficient to define the two coefficients $u$ and $v$ in $\bar\Psi^{(1)}_2(\Z) = \Z + u +v\Z^{-1}$.
In (\ref{1RPsi1}) it is enough to look at two points, and the convenient choice is $\Z=q$ and $\Z=q^{-1}$, which gives
\be
\frac{q^{-1}-1}{q-1}Q^{\frac{1}{2}} \bar \Psi^{(1)}_2(q^2) + \frac{q^3-1}{q-1}q^{-2}Q^{-\frac{1}{2}}\bar\Psi^{(1)}_2(1)
= q^{-1} (Q^{\frac{1}{2}}+Q^{-\frac{1}{2}}) \bar\Psi^{(1)}_2(q) \nn \\
\frac{q^{-3}-1}{q^{-1}-1}Q^{\frac{1}{2}} \bar \Psi^{(1)}_2(1) + \frac{q-1}{q^{-1}-1}q^{-2}Q^{-\frac{1}{2}}\bar\Psi^{(1)}_2(q^{-2})
= q^{-1} (Q^{\frac{1}{2}}+Q^{-\frac{1}{2}}) \bar\Psi^{(1)}_2(q^{-1})
\ee
Multiplying the first equation by $Q$ and subtracting the second, we eliminate the terms with $\bar\Psi^{(1)}_2(1)$.
Remarkably exactly the same procedure subtracts $\bar\Psi^{(1)}_2(q^{-1})$ from $Q\bar\Psi^{(1)}_2(q)$, which gives zero
due to the first ($j=1$) equation in (\ref{linsysPsi1}).
The remaining two terms combine into the second equation ($j=2$).

The same conspiracy persists for higher $m$.
The underlying reason is that (\ref{1RPsi1}) actually has {\it polynomial}
rather than infinite series solution for integer $m$: truncation typical for hypergeometric type functions occurs.
Technically, we substitute $\Z=q^{m+1-2k}$ with $k=1,\ldots,m$ into (\ref{1RPsi1})
and multiply it by $  Q^{m-k}$, which gives a system

{\footnotesize
\be
k=1 & \underline{\frac{q^{-1}-1}{q^{m-1}-1} Q^{m- \frac{1}{2}}\bar\Psi^{(1)}_m(q^{m})}
+  \underline{\underline{\underline{\frac{q^{m-1}-q^{-m}}{q^{m-1}-1} Q^{m-\frac{3}{2}} \bar\Psi^{(1)}_m(q^{m-2})}}}
-\underline{\underline{q^{-\frac{m}{2}}  ( Q^{m- \frac{1}{2}} + Q^{m- \frac{3}{2}})\bar\Psi^{(1)}_m(q^{m- 1})}} = 0
\nn \\
k=2 & \underline{\underline{\underline{\frac{q^{-3}-1}{q^{m-3}-1} Q^{m- \frac{3}{2}}\bar\Psi^{(1)}_m(q^{m-2})}}}
+   \frac{q^{m-3}-q^{-m}}{q^{m-3}-1}Q^{m-\frac{5}{2}} \bar\Psi^{(1)}_m(q^{m-4})
- q^{-\frac{m}{2}}  ( Q^{m- \frac{3}{2}} + Q^{m- \frac{5}{2}})\bar\Psi^{(1)}_m(q^{m-3}) = 0
\nn \\ \nn \\
& \ldots
\nn \\ \nn \\
k=m-1 & \frac{q^{3-2m}-1}{q^{3-m}-1} Q^{ \frac{3}{2}}\bar\Psi^{(1)}_m(q^{-m+4})
+ \underline{\underline{\underline{ \frac{q^{3-m}-q^{-m}}{q^{3-m}-1}Q^{ \frac{1}{2}} \bar\Psi^{(1)}_m(q^{-m+2})}}}
- q^{-\frac{m}{2}} ( Q^{ \frac{3}{2}} + Q^{ - \frac{1}{2}})\bar\Psi^{(1)}_m(q^{-m+3}) = 0
\nn \\
k=m &  \underline{\underline{\underline{\frac{q^{1-2m}-1}{q^{1-m}-1} Q^{ \frac{1}{2}}\bar\Psi^{(1)}_m(q^{-m+2})}}}
+  \underline{\frac{q^{1-m}-q^{-m}}{q^{1-m}-1}Q^{ -\frac{1}{2}} \bar\Psi^{(1)}_m(q^{-m})}
- \underline{\underline{q^{-\frac{m}{2}} ( Q^{ \frac{1}{2}} + Q^{ - \frac{1}{2}})\bar\Psi^{(1)}_m(q^{-m+1})}}  = 0
\nn
\ee
}
Clearly the last equation coincides with the first one,
the next to last with the second one and so on,
provided $Q^j \bar\Psi^{(1)}_m(q^j) =   \bar\Psi^{(1)}_m(q^{-j})$ for all $j=1,\ldots,m$,
i.e. if (\ref{linsysPsi1}) is satisfied.
Note that this coincidence happens separately term-by-term: the underlined terms are equal to each other,
the twice-underlined ones are equal to each other, etc.
There is no mixture between the neighboring lines.

\subsection{The Hamiltonian  $\hat {\bar H}^{(-1,1)}$ from the first ray   }

The Ruijsenaars Hamiltonian (\ref{HRui1}) is not the only one, which acts nicely on the BAF.
From the DIM perspective, this Hamiltonian lies on the vertical ray, while the claim of \cite{MMP1} is that the $a$-th twisted BAF
is also an eigenfunction of the Hamiltonians of the $a$-th integer ray.
In the case of $a=1$, the first Hamiltonian for the first ray is:
\be
\hat {\overline {H}}^{(-1,1)}_1(z) = {q^{-z}\over q^{1\over 2}}\frac{q^z t-1}{q^z-1} e^{\p_z}
+  {q^{z}\over q^{1\over 2}}\frac{q^z -t}{q^z-1}  e^{-\p_z}
\label{H11z}
\ee
and one has the eigenfunction equation
\be
\hat {\overline {H}}^{(-1,1)}_1(z)\left[q^{z^2\over 2}\Psi^{(1)}_m(\lambda,z)\right]=
q^{-{m\over 2}}\left(q^{\lambda\over 2}+q^{-{\lambda\over 2}}\right)\left[q^{z^2\over 2}\Psi^{(1)}_m(\lambda,z)\right]
\ee
$\hat {\overline {H}}^{(-1,1)}_1$ differs from $\hat {\overline {H}}^{(0,1)}_1=\hat {\overline {H}}^{(-1,\infty)}_1$
by additional factors   $q^{-\frac{1}{2}\mp z}$, which are compensated by  the $q^{z^2\over 2}$ twist,
which leaves $\Psi^{(1)}_m(\lambda,z)$ as a (twisted) eigenfunction with the same eigenvalue as in (\ref{1RPsi1}).

Therefore, the analysis then repeats that in the previous section.

\subsection{Warm-up example: $\bar\Psi^{(2)}_m$  }

Now we turn to a more interesting situation with $a>1$.

First of all we need to choose an $a$-generalization of (\ref{Psi1}).
It will be  \cite{MMP1}
\be
\bar\Psi_m^{(a)} =  q^{\frac{mz}{2}} \sum_{k=0}^{am} q^{-\frac{kz}{a}} \psi^{(a)}_{m,k}
\ee
This is often multiplied by $q^{\frac{\lambda z}{2a}}$, but we will separate this  factor from $\bar\Psi$.
To preserve the polynomial form of the ansatz we will work in terms of $\Z=q^{\frac{z}{a}}$ and $Q = q^{\frac{\lambda}{a}}$,
then the difference equation will always be of the first order, while the conventional $z$ is shifted by $a$,
see (\ref{H12z}) below.

We begin from the simplest case $(a,m)=(2,1)$ and look for solution of the form
\be
\bar\Psi^{(2)}_1(\Z) =  \Z \left(1+ \sum_{k=1}^{2} \Z^{-k}\psi^{(2)}_{1,k}\right)
= \Z + \psi^{(2)}_{1,1} + \frac{\psi^{(2)}_{1,2}}{\Z}
\label{Psi21}
\ee
to  Chalykh's linear system, which now consists of just two equations
\be
Q^{\frac{1}{2}} \bar\Psi^{(2)}_1(\pm q^\frac{1}{2}) = \pm Q^{-\frac{1}{2}} \bar\Psi^{(2)}_1(\pm q^{-\frac{1}{2}}),
\label{linsysPsi2m1}
\ee
where $Q=q^{\frac{\lambda}{2}}$.
The solution  is
\be
Q\left(\pm \sqrt{q} + \psi_1  \pm \frac{\psi_2}{\sqrt{q}}\right)
= \pm \left(\pm \frac{1}{\sqrt{q}} + \psi_1  \pm \psi_2\sqrt{q}\right)
\ \ \Longrightarrow \ \ \left\{\begin{array}{c}
Q\psi_1 = \frac{1}{\sqrt{q}} + \sqrt{q}\psi_2 \\
\frac{\psi_1}{Q}  = \sqrt{q} + \frac{\psi_2}{\sqrt{q}}
\end{array}\right. \ \ \Longrightarrow \ \
\begin{array}{c}
\psi^{(2)}_{1,1} = - \frac{Q(q^2-1)}{(Q^2-q)q^{\frac{1}{2}}}\\
\psi^{(2)}_{1,2} = -\frac{Q^2q-1}{Q^2-q}
\end{array}
\nn
\ee

As a direct generalization, for arbitrary $m$ we look for a more involved solution
\be
\bar\Psi^{(2)}_m(\Z) =  \Z^{m} \left(1+ \sum_{k=1}^{2m} \Z^{-k}\psi^{(2)}_{m,k}\right)
\label{Psi2}
\ee
of the system of $2m$ equations:
\be
Q^{\frac{j}{2}} \bar\Psi^{(2)}_m(\pm q^\frac{j}{2}) = \pm Q^{-\frac{j}{2}} \bar\Psi^{(2)}_m(\pm q^{-\frac{j}{2}}),
\ \ \ \ j=1,\ldots, m
\label{linsysPsi2}
\ee
In this case $(a=2)$ the system has a well-described solution, provided by eq.(\ref{psi2ans}) above.

\subsection{The first Hamiltonian for $a=2$  }

The Hamiltonian $\hat H^{(-1,2)}_1$  is already non-trivial and is beyond the familiar Ruijsenaars set.
It is:
 \be
\hat {\overline {H}}^{(-1,2)}_1(z)&= &\ {q^{{1\over 2}-z}\over q}
    \frac{(q^{z-m} - 1)}{(q^z - 1)}\frac{(q^{z+1-m} - 1)}{(q^{z+1} - 1)}
    e^{2\p_z}+ \ {q^{z-{1\over 2}}\over q}
    \frac{(q^{-z-m} - 1)}{(q^{-z} - 1)}\frac{(q^{1-z-m} - 1)}{(q^{1-z} - 1)}
    e^{-2\p_z}
    \nn \\
    & + &{q^{1\over 2}(q^{-m} - q)(q^{-m} - 1)}{q^{\frac{z}{2}}+q^{-\frac{z}{2}}\over (q^{z+1}-1)(q^{1-z}-1)}
\label{H12z}
\ee
where we substituted $t=q^{-m}$.
We need to rewrite it in terms of $\Z=q^{\frac{z}{a}}=q^{\frac{z}{2}}$
and also  multiply $\bar\Psi_m^{(2)}$ by $q^{\frac{z^2-z}{4}}q^{\frac{\lambda z}{4}}$.
Then the eigenvalue equation becomes:
\be
q^{-\frac{z^2-z}{4}}q^{\frac{-\lambda z}{4}}\hat {\overline {H}}^{(-1,2)}_1(z)   q^{\frac{z^2-z}{4}}q^{\frac{\lambda z}{4}}
\bar\Psi^{(2)}_m  =
\nn
\ee
\be
= \frac{1}{q^{\frac{1}{2}}\Z^2} \frac{(\Z^2q^{-m}-1)(\Z^2q^{1-m}-1)}{(\Z^2-1)(q\Z^2-1)} q^{\frac{1}{2}}\Z^2Q \bar\Psi^{(2)}_m(q\Z)
+ \frac{\Z^2}{q^{\frac{3}{2} }}\frac{(\Z^2q^m-1)(\Z^2q^{m-1}-1)}{(\Z^2-1)(\Z^2-q)} q^{1-2m} \frac{q^{\frac{3}{2}}}{\Z^2Q} \bar\Psi^{(2)}_m(q^{-1}\Z)
+ \nn \\
-  q^{\frac{1}{2}-2m}(q^m-1)(q^{m+1}-1)\frac{\Z(\Z^2+1)}{(\Z^2q-1)(\Z^2-q)} \bar\Psi^{(2)}_m(\Z) =
\nn
\ee
\be
= \frac{(\Z^2q^{-m}-1)(\Z^2q^{1-m}-1)}{(\Z^2-1)(q\Z^2-1)}  Q \bar\Psi^{(2)}_m(q\Z)
+    \frac{(\Z^2-q^{-m})(\Z^2-q^{1-m})}{(\Z^2-1)(\Z^2-q)} Q^{-1}  \bar\Psi^{(2)}_m(q^{-1}\Z)
+ \nn \\
-  q^{\frac{1}{2} }(1-q^{-m})(q-q^{-m})\frac{\Z(\Z^2+1)}{(\Z^2q-1)(\Z^2-q)} \bar\Psi^{(2)}_m(\Z)
\ \ \ =\ \ \  q^{-m}(Q+Q^{-1})\bar\Psi^{(2)}_m(\Z)
\label{H2action}
\ee
Note that the double shift $e^{2\p_z}$ multiplies $\Z=q^{\frac{z}{2}}$ by the first power of $q$,
and converts $q^{\frac{z^2-z}{4}}q^{\frac{\lambda z}{4}}$ into additional factor $q^{\frac{1}{2}}Q\Z^2$.
Likewise $e^{-2\p_z}$ divides $\Z=q^{\frac{z}{2}}$ by the first power of $q$,
and converts $q^{\frac{z^2-z}{4}}q^{\frac{\lambda z}{4}}$ into additional factor $\frac{q^{\frac{3}{2}}}{Q\Z^2}$.

And, indeed, \eqref{psi2ans} does satisfy this eigenfunction equation.

\subsection{A regular approach to constructing Hamiltonians}

The apparent problem with the Hamiltonian (\ref{H12z}) is that it is not obvious to guess and to generalize,
both to higher $N$ and to higher $a$.
A regular approach was explained in \cite{MMP} for the generic Hamiltonians
{\it not} restricted to the locus $\sum_{i=1}^N z_i =0$.
Then the counterparts of (\ref{H11z}) and (\ref{H12z}) are:
\be
\hat {  {\cal H}}^{(-1,1)}(\vec x) =  \sum_i  \underline{\frac{1}{q^{\frac{1}{2}}x_i}}
\left(\prod_{j\neq i}\frac{x_it -x_j}{x_i-x_j}\right) e^{\p_{z_i}}
\label{calH11}
\ee
and
\be
\hat {  {\cal H}}^{(-1,2)}(\vec x) =
\sum_i \left(\prod_{j\neq i}\frac{(x_it-x_j)(x_iqt-x_j)}{(x_i-x_j)(qx_i-x_j)}\right)
\underline{\frac{1}{q x_i}}e^{2\p_{z_i}} +
\nn \\
+  (t - q)(t - 1)
    \sum_{i < j} \left(\prod_{k \neq i,j}
    \frac{(t x_i - x_k)(t x_j - x_k)}{(x_i - x_k)(x_j - x_k)}\right)
    \frac{x_i+x_j}{(q x_i - x_j)(q x_j - x_i)}\, e^{\p_{z_i}+\p_{z_j}}
\label{calH12}
\ee
and for $N=2$ they act on
\be
Q^{\frac{z_1-z_2}{2}} \underline{q^{\frac{z^2_1+z_2^2}{2a}}}
\cdot \Psi^{(a)}_m(z_1,z_2)
\ee
Here $Q=q^{\frac{\lambda}{a}}$ and
\be
\Psi^{(a)}_m(z_1,z_2) =
q^{\frac{ma}{2}\frac{z_1-z_2}{a}}\left(1 + \sum_{k=1}^{ma} q^{-\frac{k(z_1-z_2)}{a}}\psi^{(a)}_{m,k}\right)
\ee
Note that the shift operators $e^{\p_{z_i}}$ in (\ref{calH11}) and (\ref{calH12}) act on $z_i$, while the corresponding $x$-variables are not shifted,
but instead are multiplied by $q$: therefore in \cite{MMP1} we used the notation $e^{\p_{z_i}}=q^{\hat D_i}$
with dilatation operator $\hat D_i = x_i\frac{\p}{\p x_i}$.
The role of the strange underlined exponential in the prefactor is to cancel
the equally strange underlined denominators in (\ref{calH12}) to convert them for $N=2$
and $t=q^{-m}$ into
\be
q^{-\frac{z^2_1+z_2^2}{2}} Q^{-\frac{z_1-z_2}{2}}\
\hat {  {\cal H}}^{(-1,1)}(\vec x)\ Q^{\frac{z_1-z_2}{2}}q^{\frac{z^2_1+z_2^2}{2}}\Psi^{(1)}_m(z_1,z_2)=
\nn \\
= \left( \frac{q^{z_1-m}-q^{z_2} }{q^{z_1}-q^{z_2}} Q^{\frac{1}{2}}e^{\p_{z_1}}
+ \frac{q^{z_2-m}-q^{z_1} }{q^{z_2}-q^{z_1}} Q^{-\frac{1}{2}}e^{\p_{z_2}}\right)\Psi^{(1)}_m(z_1,z_2)
\ \ \  = \ \ \ q^{-\frac{m}{2}}(Q+Q^{-1}) \ \ \ \Psi^{(1)}_m(z_1,z_2)
\ee
and
\be
q^{-\frac{z^2_1+z_2^2}{4}} Q^{-\frac{z_1-z_2}{2}}\
\hat {  {\cal H}}^{(-1,2)}(\vec x)\ Q^{\frac{z_1-z_2}{2}}q^{\frac{z^2_1+z_2^2}{4}}\Psi^{(2)}_m(z_1,z_2)=
\nn \\
= \left(  \frac{(q^{z_1-m}-q^{z_2})(q^{z_1+1-m}-q^{z_2})}{(q^{z_1}-q^{z_2})(q^{z_1+1}-q^{z_2})}
Qe^{2\p_{z_1}} +
 \frac{(q^{z_2-m}-q^{z_1})(q^{z_2+1-m}-q^{z_1})}{(q^{z_2}-q^{z_1})(q^{z_2+1}-q^{z_1})}
Q^{-1}e^{2\p_{z_2}}
+ \right. \nn \\ \left.
+   \underline{q^{\frac{z_1+z_2+1}{2}}}(q^{-m}-q)(q^{-m}-1)\frac{q^{z_1}+q^{z_2}}{(q^{z_1+1}-q^{z_2})(q^{z_2+1}-q^{z_1})}
e^{\p_{z_1}+\p_{z_2}}\right)
\Psi^{(2)}_m(z_1,z_2)
= \nn \\
 = q^{-m}(Q+Q^{-1})\Psi^{(2)}_m(z_1,z_2)
\label{H12new}
\ee
Note that this is very similar to (\ref{H2action}), still different in details.
Also the ugly underlined factors (\ref{calH12}) are traded for an equally ugly underlined factor in (\ref{H12new}).

The advantage  of (\ref{calH12}) over (\ref{H11z}) and (\ref{H12z})
is that  Hamiltonians (\ref{calH12}) are {\it canonical}:
there is a generalizable procedure, which allows one to build them.
Namely, they are obtained by a simple iteration
\be \label{eq:ham-repeated-commutator}
\hat {\cal H}_1^{(-1,a+1)}=[\hat {\cal H}_1^{(-1,a)},\hat {\cal H}_{MR}]
\ee
from
\be \label{eq:hamiltonian-sum-xs}
{\cal H}_1^{(-1,0)}=\sum_{i=1}^Nx_i^{-1}
\ee
and
\be \label{eq:macdonald-ruijsenaars-hamiltonian}
\hat {\cal H}_{MR}={\sqrt{q}\over q-1}\ \sum_{i=1}^N\prod_{j\ne i}{tx_i-x_j\over x_i-x_j}e^{\p_{z_i}}
\ee

With the help of this procedure we can easily construct the Hamiltonian for the third ray:
\be
\hat{\cal H}^{(-1,3)}_1 = \sum_i \left(\prod_{j\neq i} \frac{(x_it-x_j)(x_iqt-x_j)(x_iq^2t-x_j)}{(x_i-x_j)(qx_i-x_j)(q^2x_i-x_j)}\right)
\underline{\frac{1}{q^{\frac{3}{2}}x_i}}\,e^{3\p_{z_i}}  + \nn  \\
+ \frac{(t-q)(t-1)}{q^{\frac{1}{2}}}\sum_{i< j}
\left(\prod_{k\neq i,j}\frac{(tx_i-x_k)(tx_j-x_k)(tqx_i-x_k)}{(x_i-x_k)(qx_i-x_k)(x_j-x_k)}\right)
\frac{(tx_i-x_j)\Big(q(x_i+x_j)+x_j\Big)}{(x_i-x_j)(qx_j-x_i)(q^2x_i-x_j)} e^{2\p_{z_i}+\p_{ z_j}} +\\
+ q^{\frac{1}{2}}(t-q)^2(t-1)^2 \sum_{i< j< k} \left(\prod_{m\ne i,j,k}{(tx_i-x_m)(tx_j-x_m)(tx_j-x_m)\over (x_i-x_m)(x_j-x_m)(x_j-x_m)}
\right) \times\nn\\
\times\frac{(x_ix_j+x_ix_k+x_jx_k)P_3(x_i,x_j,x_k)}{(qx_i-x_j)(qx_i-x_k)(qx_j-x_i)(qx_j-x_k)(qx_k-x_i)(qx_k-x_j)}
e^{\p_{z_i}+\p_{z_j}+\p_{z_k}}
 \nn
\ee
with $P_3(x_i,x_j,x_k):=(t^2q+t^2+tq^2+t+q^2+q) x_ix_jx_k
- tq(x_i^2x_j+x_i^2x_k+x_j^2x_i+x_j^2x_k+x_k^2x_i+x_k^2x_j)$. Introducing the operator
\be
{\cal D}_i:=\prod_{j\ne i}{tx_i-x_j\over x_i-x_j}e^{\p_{z_i}}
\ee
one can rewrite this Hamiltonian as
\be
\hat{\cal H}^{(-1,3)}_1 =\sum_i \underline{\frac{1}{q^{\frac{3}{2}}x_i}}\,{\cal D}_i^3+
\frac{(t-q)(t-1)}{q^{\frac{1}{2}}}\sum_{i< j}
\frac{(tx_i-x_j)\Big(q(x_i+x_j)+x_j\Big)}{(x_i-x_j)(qx_j-x_i)(q^2x_i-x_j)} {\cal D}_i^2{\cal D}_j +\\
+ q^{\frac{1}{2}}(t-q)^2(t-1)^2 \sum_{i< j< k} \underbrace{\frac{(x_ix_j+x_ix_k+x_jx_k)P_3(x_i,x_j,x_k)}{(qx_i-x_j)(qx_i-x_k)(qx_j-x_i)(qx_j-x_k)(qx_k-x_i)(qx_k-x_j)}}_{{\alpha_1 x_i+\alpha_2(x_j+x_k)\over (qx_i-x_j)(qx_i-x_k)}+{\alpha_1 x_j+\alpha_2(x_i+x_k)\over (qx_j-x_i)(qx_j-x_k)}+{\alpha_1 x_k+\alpha_2(x_i+x_j)\over (qx_k-x_i)(qx_k-x_j)}}
{\cal D}_i{\cal D}_j{\cal D}_k
\ee
where $\alpha_1=q{q^2t(q+t)+(q^2+t-1)(q-t)\over (q-1)(5q^2+2q^3+1)}$, $\alpha_2=-{q+1\over 2}{(2q^2t+1)(q+t)+(q-t-qt)(q-t)\over (q-1)(5q^2+2q^3+1)}$.

\bigskip

For $N=2$ we get from this expression the $a=3$ analogue of (\ref{H12new}):
\be
q^{-\frac{z^2_1+z_2^2}{6}} Q^{-\frac{z_1-z_2}{2}}\
\hat {  {\cal H}}^{(-1,3)}(\vec x)\ Q^{\frac{z_1-z_2}{2}}q^{\frac{z^2_1+z_2^2}{6}}\Psi^{(3)}_m(z_1,z_2)=
\nn
\ee
\be
\!\!\!\!\!\!\!\!\!\!\!\!\!
= \left\{  \frac{(q^{z_1-m}-q^{z_2})(q^{z_1+1-m}-q^{z_2})(q^{z_1+2-m}-q^{z_2})}{(q^{z_1}-q^{z_2})(q^{z_1+1}-q^{z_2})(q^{z_1+2}-q^{z_2})}
Q^{\frac{3}{2}} e^{3\p_{z_1}} +
 \frac{(q^{z_2-m}-q^{z_1})(q^{z_2+1-m}-q^{z_1})(q^{z_2+2-m}-q^{z_1})}{(q^{z_2}-q^{z_1})(q^{z_2+1}-q^{z_1})(q^{z_2+2}-q^{z_1})}
Q^{-\frac{3}{2}}e^{3\p_{z_2}}
+ \right. \nn \\ \left.
+   q^{\frac{1}{3}}(q^{-m}-q)(q^{-m}-1)
\left(q^{\frac{2z_1+z_2}{3}}(q^{z_1-m}-q^{z_2})\frac{q^{z_1+1}+q^{z_2+1}+q^{z_2}}{(q^{z_1+2}-q^{z_2})(q^{z_1}-q^{z_2})(q^{z_2+1}-q^{z_1})}
e^{2\p_{z_1}+\p_{z_2}}
+\right.\right. \nn \\ \left.\left.
+q^{\frac{z_1+2z_2}{3}}(q^{z_2-m}-q^{z_1})\frac{q^{z_2+1}+q^{z_1+1}+q^{z_1}}{(q^{z_2+2}-q^{z_1})(q^{z_2}-q^{z_1})(q^{z_1+1}-q^{z_2})}
e^{\p_{z_1}+2\p_{z_2}}
\right)
\right\}
\Psi^{(3)}_m(z_1,z_2)
= \nn
\ee
\be
 = q^{-\frac{3m}{2}}(Q^{\frac{3}{2}}+Q^{-\frac{3}{2}})\Psi^{(3)}_m(z_1,z_2)
\label{H12new3}
\ee
Generalizations to higher $N$ and $a$ are straightforward,
but somewhat lengthy to write down explicitly, and in practical calculations
the iterative description \eqref{eq:ham-repeated-commutator}-\eqref{eq:macdonald-ruijsenaars-hamiltonian} is sufficient.

Note that the (Hamiltonian) reduction of thus obtained Hamiltonians to the center-of-mass
variable $z = z_1 - z_2$ is rather straightforward, but one can equally well repeat
the iteration \eqref{eq:ham-repeated-commutator} starting from
center-of-mass counterparts of
\eqref{eq:hamiltonian-sum-xs} and \eqref{eq:macdonald-ruijsenaars-hamiltonian}.

\section{Conclusion
\label{conc}}

The main goal of this note is to illustrate the content of \cite{MMP1} by explicit examples
of {\bf what the solutions to simple Chalykh's equations for BAF's look like}.
We see that for $N>2$ these solutions are originally somewhat ugly,
but get {\bf decomposed into nice factorized expressions}.
These expression are somewhat difficult to generalize,
moreover it is unclear if factorized decomposition is unique,
and what kind of additional symmetry it reflects.
It appears that important is the condition that extra factors depend only on
variables associated to non-simple roots, though at the moment it is not clear how to put
this in practice beyond the case of $N=3$ (where there is just one non-simple root,
 which delivers an extra degree of control).

A complementary goal is to advertise the twisted BAF's: the claim is that they are
\textbf{not more complicated} to work with than their usual non-twisted counterparts.
At the same time, this twisted case is much more interesting for applications to the new
sets of coexisting integrable systems \cite{MMP} implied by the q,t-deformation \cite{Ch3} of the WLZZ models \cite{China,Ch12}
and by the DIM algebra \cite{DIM}.
In this case, the Chalykh's equations are modified just slightly, but the solutions get far more involved
and do not factorize immediately even at $N=2$.
They are no longer related to the Macdonald polynomials, at least directly,
which opens a set of questions about their meaning.
Again, in this paper, we just raise and demonstrate the problem, postponing possible resolutions for the future.
What we do, however, is provide additional evidence that {\bf the twisted BAF's are indeed the eigenfunctions
of the Hamiltonians for non-trivial rays}, at least for the integer rays.
Generalization to rational rays is still an open problem.

Essentially, this whole text is about solutions of linear equations,
i.e. a kind of Cramer's rule application to a simply looking system (\ref{linsys0})+(\ref{Psideco}). It is just amazing to see how sophisticated the outcome is, and what a variety of structures emerge seemingly out of ``nothing''.
It would be nice to find a better understanding of this phenomenon, and a better way
to present it.

\section*{Acknowledgements}

Our work is supported by RSF Grant No. 24-12-00178.

\end{document}